\documentclass[doublespacing]{elsart}

\usepackage[dvips]{graphicx}

\usepackage{amssymb}
\usepackage{amsmath}
\usepackage[linktocpage]{hyperref}

\newlength{\swidth}%
\begin{document}

\begin{frontmatter}



\title{Application of the time-dependent charge asymmetry method for longitudinal position determination in prototype proportional chambers for the $\overline{\mbox{\sf P}}${\sf ANDA} experiment.}

\author[FZ_Juelich]{Andrey Sokolov\corauthref{cor}},
\ead{a.sokolov@fz-juelich.de}
\author[FZ_Juelich]{James Ritman},
\author[FZ_Juelich]{Peter Wintz}
\corauth[cor]{Corresponding author}

\address[FZ_Juelich]{Institute f\"ur Kernphysik, Forschungszentrum J\"ulich, 52425 J\"ulich, Germany}

\begin{abstract}
The $\overline{\mbox{\sf P}}${\sf ANDA} collaboration intends to
build a state-of-the-art detector to study the physics of antiproton
annihilation in the charm mass region at the future {\sf FAIR}
facility at {\sf GSI}, Darmstadt. One major part of the
$\overline{\mbox{\sf P}}${\sf ANDA} detector is the straw tube
tracker. It will consist of about 6000 individual straws grouped in
11 double layers and filled with an $Ar$+10\,\%$\;CO_2$ gas mixture.
The required radial spatial resolution is about $150\,\mu$m. Two
different methods are considered for longitudinal coordinate
measurements - skewed double layers and a novel method based on the
time-dependent charge asymmetry. The latter method is presented in
this article.

\end{abstract}

\begin{keyword}
PANDA, FAIR, straw tube, charge division.
\PACS 25.43.+t \sep 29.40.Cs \sep 29.40.Gx
\end{keyword}
\end{frontmatter}

\section{Introduction}
\label{intro}

A new state-of-the-art general-purpose detector is proposed by the
$\overline{\mbox{\sf P}}${\sf ANDA} collaboration~\cite{Panda}. It
will be installed at the High Energy Storage Ring, {\sf HESR}, one of
the main components of the international {\sf FAIR} facility at GSI,
Darmstadt. The detector is designed to take advantage of the
extraordinary physics potential which becomes available by utilizing
high-intensity, phase-space-cooled antiproton beams with momenta
between 1 and 15 GeV/c.

\begin{figure}[htb]
   \begin{center}
      \includegraphics[width=1.2\swidth]{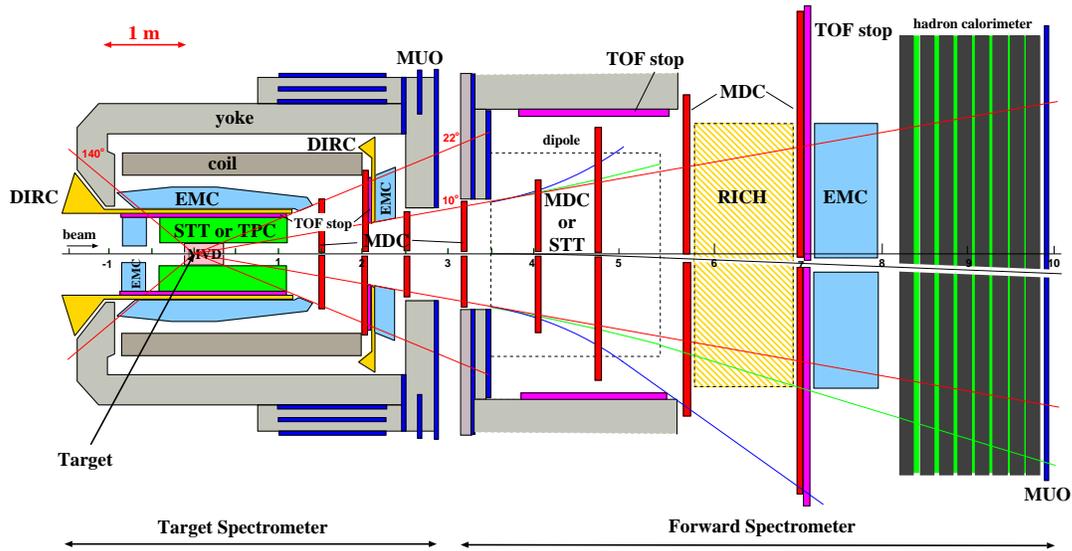}
      \caption[Sketch of the $\overline{\mbox{\sf P}}${\sf ANDA} detector from the top]{Sketch of the $\overline{\mbox{\sf P}}${\sf ANDA} detector from the top. The total length of the detector will be about 12m. Figure taken from~\cite{Panda}.}
      \label{fig:panda}
   \end{center}
\end{figure}

Due to the fixed target kinematics, the detector has an asymmetric shape and consist of two parts - the target spectrometer and the forward spectrometer, see~\autoref{fig:panda}.

The Central Tracker of the Target Spectrometer consists of two main parts: the
Micro-Vertex Detector (MVD) and the Straw Tube Tracker (STT).

\section{Straw Tube Tracker}
\label{stt} The STT consists of 11 cylindrical double layers of
straws at radial distances of between 15$\,$cm and 42$\,$cm from the
beam and with an overall length of 150$\,$cm\footnote{The exact
parameters of the STT are under investigation. For more information
see~\cite{Panda}.}. The geometrical center of the STT is shifted
35\,cm downstream with respect to the target position to provide
better acceptance in the forward hemisphere.

The individual straw tubes have diameters of 6\,mm for the five
innermost double layers, and 8\,mm for the outer layers. The gas
mixture chosen for the straw tracker is $Ar$+10\,\%$\;CO_2$. The
relatively small tube diameters and the chosen gas provide maximum
drift times comparable with the average time between events
(100\,ns).

Every other double layer can be skewed with respect to the beam axis by an angle ranging from
$2^{\circ}$ to $3^{\circ}$ to obtain
the coordinate of the track along the beam direction with about
4\,mm precision. However, the skew angle causes the stereolayers to
form a hyperboloidal shape, which makes the track reconstruction and
mechanical design more complicated. As an alternative, it is
proposed to use the charge-division technique instead of skewed
layers, thereby simplifying the overall STT design and allowing the
straws to be glued together thus forming a self-supporting
construction, and significantly reducing the amount of passive
material needed for structural reasons.

\begin{figure}[htb]
   \begin{center}
      \includegraphics[width=\swidth]{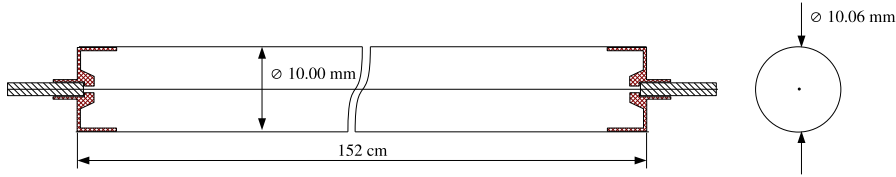}
      \caption[Longitudinal cross section of the 1.5m straw tube.]{Longitudinal (left) and transverse (right) cross sections of a 1.52-m-long straw tube.}
      \label{fig:straw}
   \end{center}
\end{figure}

The straw tubes used for testing the charge division principle are
made of two thin Mylar$^{\copyright}$ strips $12\,\mu$m in thickness
and 16\,mm in width that are wound helically with a half-overlap and
glued together. Including 6\,$\mu$m glue, the total straw tube wall thickness is $30\,\mu$m. 
The straw tubes were supplied by an industrial
manufacturer~\cite{Lamina}. The inner part of the inner strip is
metallized and the inner diameter of the straw tube is 10\,mm. The anode wire is made of
W/Re alloy and is 20$\,\mu$m in diameter. The longitudinal and transverse cross
sections of the straw tube are shown in~\autoref{fig:straw}.

\begin{figure}[htb]
   \begin{center}
      \includegraphics[width=\swidth]{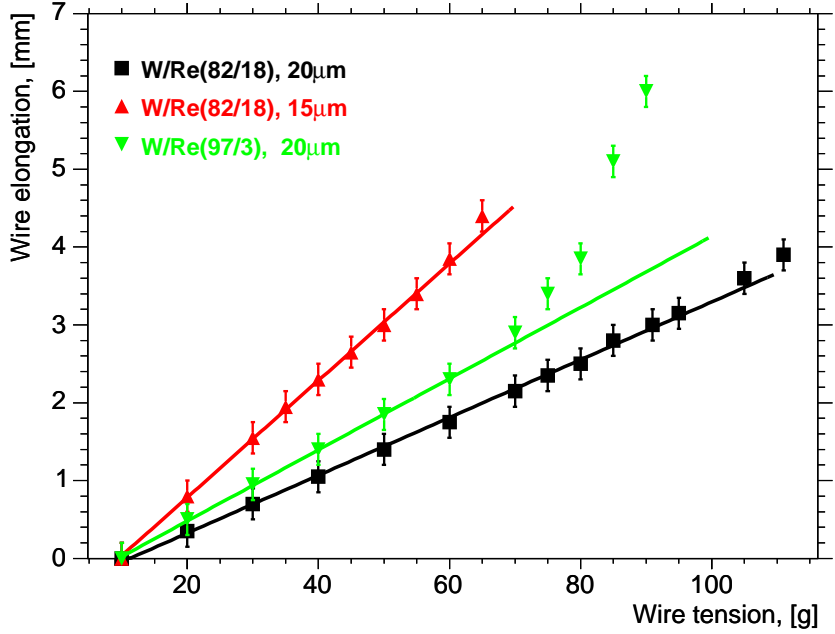}
      \caption[Wire elongation depending on applied tension.]{Anode wire elongation due to applied tension is shown for different wire diameters and Re concentrations. The length of each wire was 1\,m.}
      \label{fig:wire_tension}
   \end{center}
\end{figure}

Wires with different Re concentrations were tested. As can be seen
in~\autoref{fig:wire_tension}, wires with a higher percentage of Re
have better mechanical properties.  The region of plastic
deformation for the 20-$\mu$m diameter wire with 3\% Re already
starts at a tension of about 65\,g, while for wire of the same
diameter with 18\% Re no plastic deformation is observed up to a
wire tension of 110\,g. For the 1.52-m-long straw the required anode
wire tensions are 32\,g and 58\,g for the 15- and 20-$\mu$m diameter
wires, respectively, in order to achieve a gravitational sag of the
wires of less than 30$\,\mu$m. It is preferable to make use of wires
with an increased percentage of Re in order to have an adequate safety margin.

\section{Test Setup}
\label{setup}

\begin{figure}[htb]
   \begin{center}
      \includegraphics[width=\swidth]{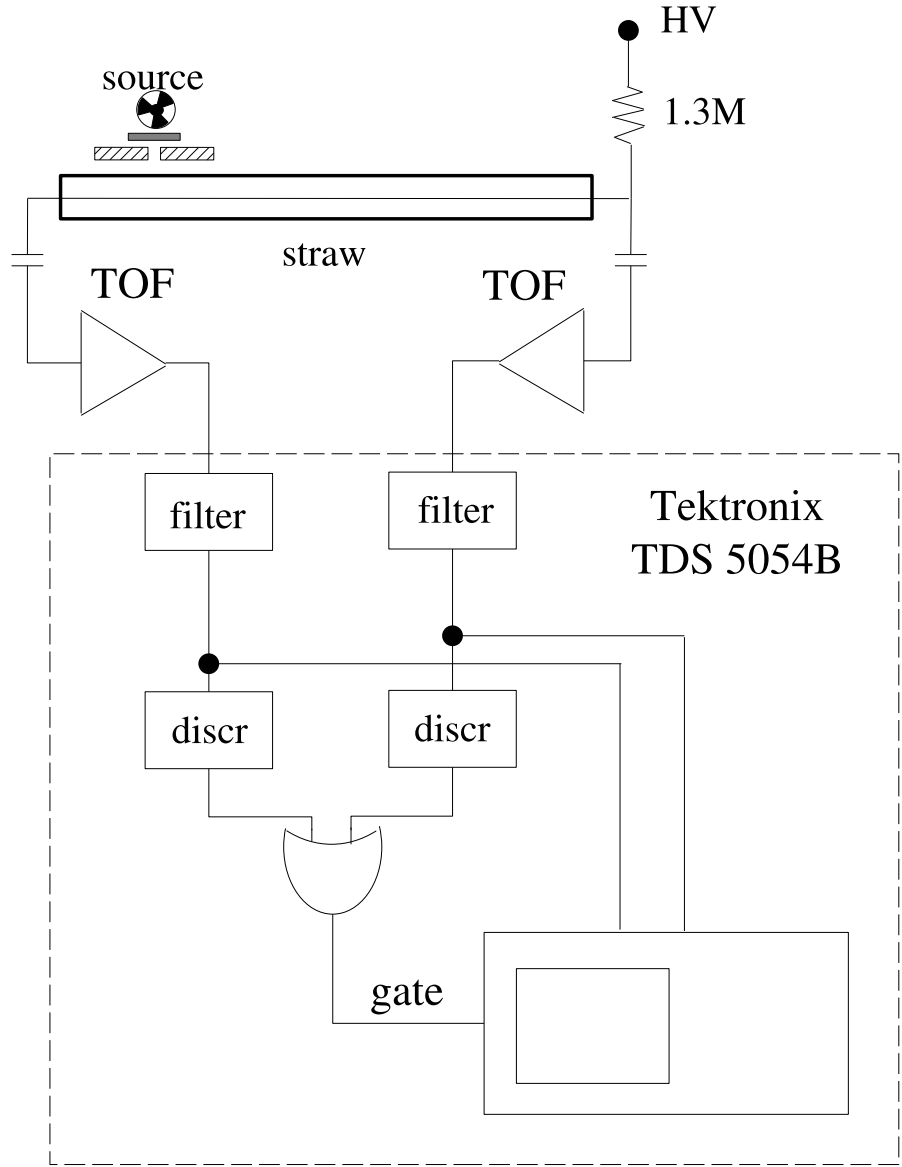}
      \caption[Experimental setup for the charge division measurement.]{Experimental layout for the charge division measurement.}
      \label{fig:test_setup}
   \end{center}
\end{figure}

The straw tube performance was studied using collimated radioactive
sources. An $^{55}$Fe source was used
to study the ultimate spatial resolution limitations of the method and a $^{90}$Sr source was used to test the charge division
performance with realistic detector response. The experimental setup is shown schematically
in~\autoref{fig:test_setup}.

\begin{figure}[htb]
   \begin{center}
      \includegraphics[width=\swidth]{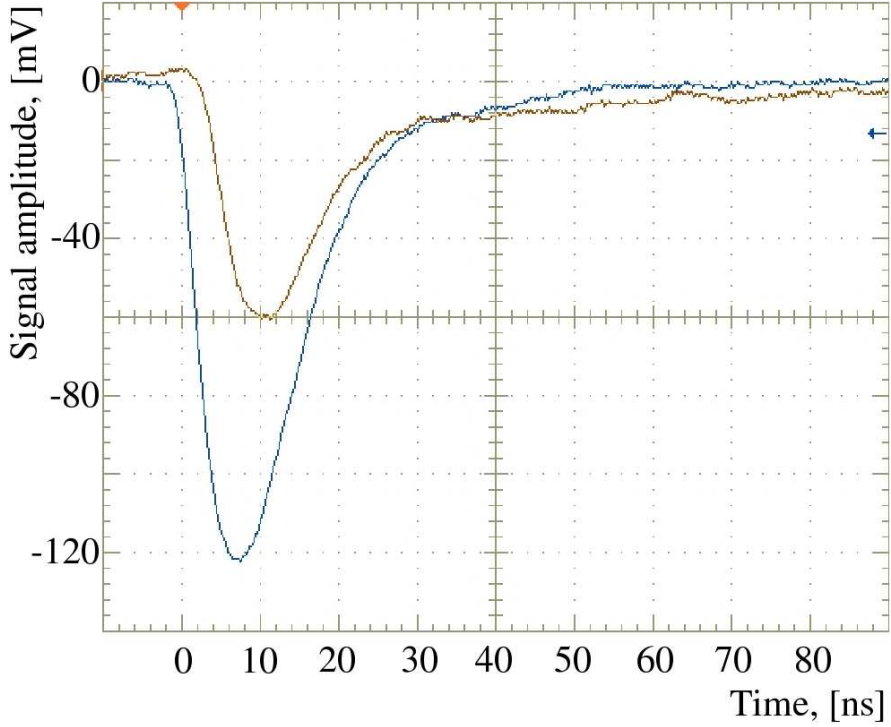}
      \caption[Comparison of the signal amplitudes from both ends of the 1.5\,m long straw tube]{Comparison of the signal amplitudes from opposite ends of the 1.52-m-long straw tube. The wire resistance is  $600\,\Omega$, and the $^{55}$Fe source is placed at the end of the straw tube.}
      \label{fig:signals}
   \end{center}
\end{figure}

The signals at the both ends of the straw are read out by COSY-TOF experiment~\cite{Tof} type preamplifiers, attached directly
to the anode wire feedthroughs. The amplifier outputs were connected
to the Tektronix$^{\copyright}$ 5054B oscilloscope, which was
used for the data acquisition. In order to cut high-frequency
noise the bandwidth of the oscilloscope was reduced to 20MHz, thus increasing
the signal rise time from 3-4\,ns to 10-15\,ns
(see~\autoref{fig:signals}). The pulse decay time was about 40\,ns.
These preamplifiers are optimized for the drift time measurement and it can therefore be expected that optimizing the shaping characteristics of the preamplifiers should further improve the results presented below.

These signals were parametrized by the following equation
\begin{equation}
   V(t) = A e^{B \ln(t) - Ct}(1+Dt+Et^2),
\label{equ:pss:res:sig}
\end{equation}
where the parameter $A$ is a scaling factor and the values of the other parameters were obtained from a fit to measured signals: $B=3.95$, $C=0.228$, $D=-3.839\cdot10^{-2}$ and $E=1.148\cdot10^{-3}$.
%

\section{Linear Charge Asymmetry}
\label{asymm}

The {\it linear charge asymmetry} due to charge sharing between the tube ends can be expressed as follows:
\begin{equation}
   \alpha(z_{true})=\frac{Q_r(z_{true})-Q_l(z_{true})}{Q_r(z_{true})+Q_l(z_{true})},
   \label{equ:assym}
\end{equation}
where $z_{true}$ is the actual position of the source along the straw tube with
$-l/2 \leqslant z_{true} \leqslant l/2$, $l$ is the real tube length,
$Q_l(z_{true})$ and $Q_r(z_{true})$ are the charges collected on the
left and right ends of the straw, respectively.

The {\it effective} tube length $l_{eff}$ can be defined as follows:
\begin{equation}
      l_{eff} = \: \frac{l}{\alpha(l/2)}.
   \label{equ:length}
\end{equation}

Finally, the reconstructed longitudinal coordinate $z_{rec}$ can be calculated in the following way:
\begin{equation}
   z_{rec}=\frac{l_{eff}}{2}\: \alpha(z_{true}).
   \label{equ:coor}
\end{equation}
The actual values of $Q_r(l/2)$ and $Q_l(l/2)$ depend on the ratio
of the wire resistance to the input impedance of the preamplifiers, which were determined from calibration measurements.

\begin{figure}[htb]
   \begin{center}
      \includegraphics[width=\swidth]{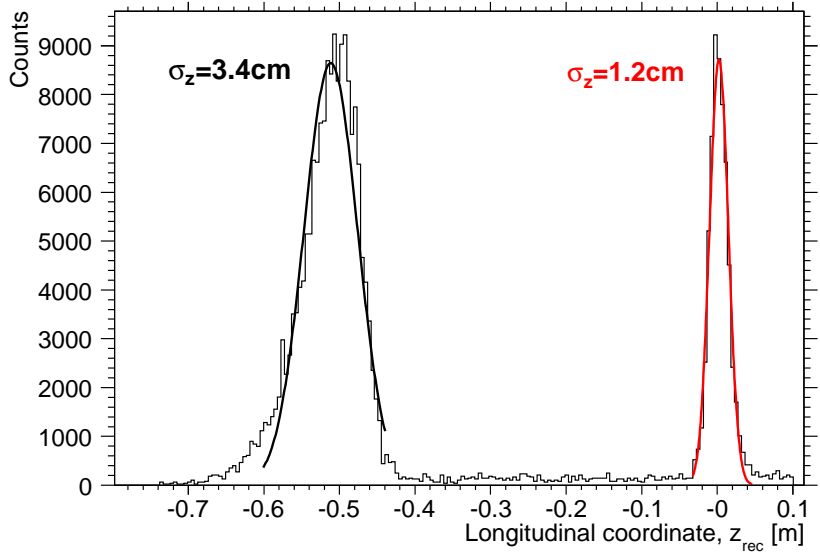}
      \caption[$z$-resolution for a $1\,m$ straw.]{Reconstructed longitudinal position distribution for an $^{55}$Fe source located at $z_{true}$=-51.0\,cm and 0.0\,cm along the $1.05$-m-long straw tube with the W/Re (97/3) wire of 20-$\mu$m diameter. The position resolution at the center and at the end was found to be $\sigma=1.2\pm0.2_{sys}\,$cm and $\sigma=3.4\pm0.2_{sys}\,$cm, respectively. The x-scale corresponds to one half of the tube length.}
      \label{fig:short_straw}
   \end{center}
\end{figure}

It should be noted that the definition of $l_{eff}$ implicitly
assumes that the input impedances of the preamplifiers and
their relative gain are identical. To determine the amplifier characteristics, the
collimated $^{55}$Fe source was placed exactly at the center (i.e. $z_{true}=0$)
of the 1.05-m-long straw tube with the anode wire made of W/Re (97/3)
alloy with a resistance of $260\pm10\,\Omega/m$. The
fitted mean value of the reconstructed position distribution was found to be
$z_{rec}=\,1.23\,$cm compared to the expected value $z_{true}=0$. After correcting
the input impedances and gains of the preamplifiers, the mean value was
reduced to $z_{rec}=2\,$mm, as shown
by the right peak in~\autoref{fig:short_straw}. The source was then placed at
$z_{true}=\,-51.0\,$cm with respect to the tube center. The
reconstructed source position was $z_{rec}=\,-51.2\,$cm. The maximum observed
deviation of $z_{true}-z_{rec}$ was 2\,mm, which is significantly smaller
than the typical spatial resolution of about 1 to 3\,cm
(cf.~\autoref{fig:short_straw}). This small deviation from the
linear behavior is included as a systematic error below.
Therefore, the longitudinal resolution (defined as the standard deviation of the Gaussian fit to the measured position distribution) is determined to be
$\sigma=1.2\pm0.2_{sys}\,$cm or $1.2$\,\% of the length in the center and $\sigma=3.4\pm0.2_{sys}\,$cm or
$3.2$\,\% at the end of the tube.

\begin{figure}[htb]
   \begin{center}
      \includegraphics[width=\swidth]{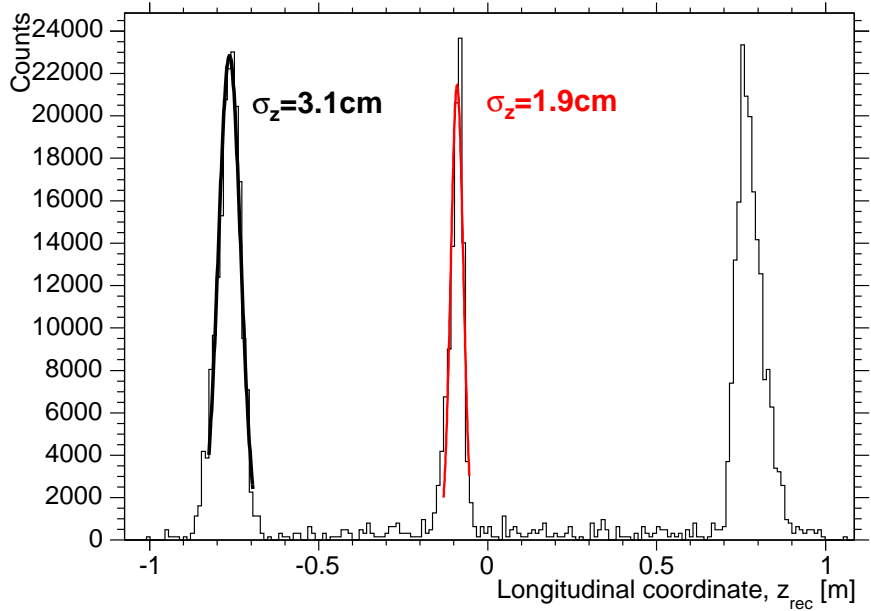}
      \caption[The $z$-resolution for the $1.52\,m$ straw tubes.]{Reconstructed longitudinal position distribution for an $^{55}$Fe source located at $z_{true}=\,\pm76.0$\,cm and -8.0\,cm along the $1.52$-m-long straw tube with the W/Re (82/18) wire of 20\,$\mu$m diameter. The position resolutions near the center and at the ends were found to be $\sigma=1.9\pm0.7_{sys}\,$cm and $\sigma=3.1\pm0.7_{sys}$\,cm, respectively.}
      \label{fig:long_straw}
   \end{center}
\end{figure}

A 1.52-m-long straw tube with an anode wire made of W/Re (82/18) alloy
was used for the further tests. This wire has a specific resistance of $400\pm15\,\Omega/m$.
The total resistance of the 1.52-m wire
was 600$\pm20\,\Omega$. The increased resistance leads to a higher
signal asymmetry and thus to a better relative spatial resolution. 
The $^{55}$Fe source was placed at three locations: $z_{true}=\,\pm76.0\,$cm and -8.0\,cm.
 \autoref{fig:long_straw} shows the $z_{rec}$ spectrum after applying the calibration procedure.
Longitudinal resolutions of $\sigma=1.9\pm0.7_{sys}\,$cm (1.3\%) and $\sigma=3.1\pm0.7_{sys}$\,cm 
(2.1\%) were determined near the tube center and at the end, respectively.

When using the charge division technique, the longitudinal spatial resolution improves when the signal asymmetry is increased because the resolution is dominated
by noise. The noise can be separated
into series and parallel components~\cite{pullia}. 
Parallel-noise components, such as RF-pickup or noisy power supplies, induce correlated signals of the same sign at both ends of 
the straw tube. Series noise induces either anticorrelated or uncorrelated signals in the preamplifiers, e.g. 
thermal noises in the anode wire and preamplifiers, respectively.
Pure incoherent series noise will result in a constant spatial resolution along the anode wire. Pure anticorrelated series noise will have the poorest resolution in the center of the wire, and pure parallel noise will have the best resolution in the center of the wire.

\section{Time-Dependent Charge Asymmetry}
\label{time}

The {\it time-dependent charge asymmetry} is a novel method to determine the longitudinal coordinate, and it differs from the {\it linear
charge asymmetry} by taking into account not only the signal charges but
also the differences in the arrival times of the signals, allowing for a
significant increase of the asymmetry magnitude up to the maximum value of 1.

\begin{figure}[htb]
   \begin{center}
      \includegraphics[width=\swidth]{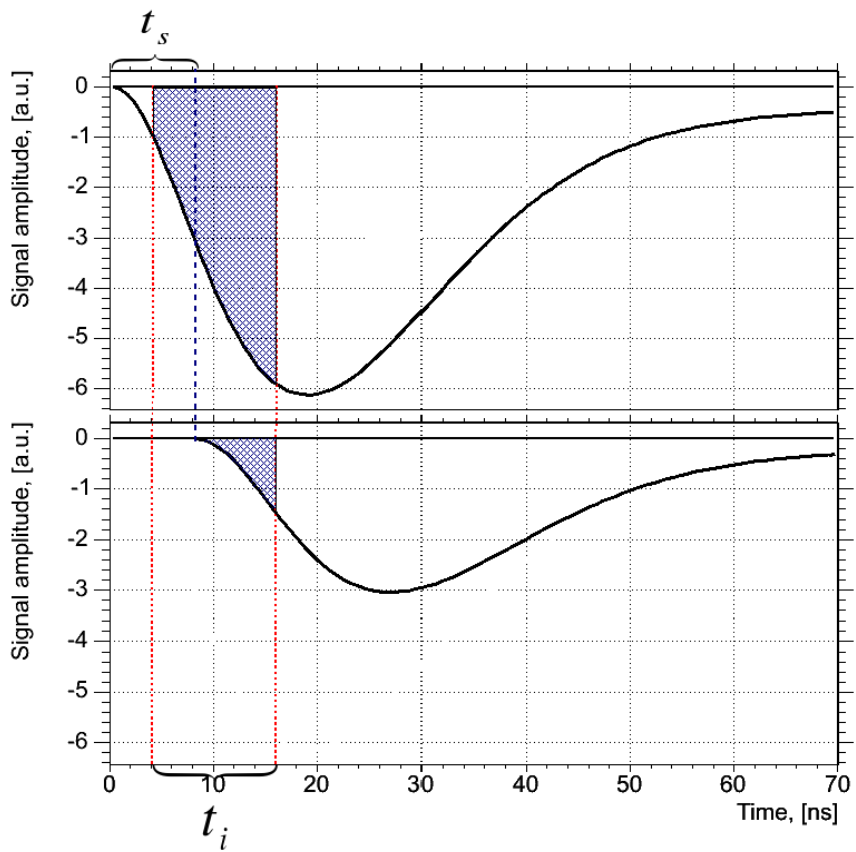}
      \caption[Time shift between signals from opposite ends of the straw tube.]{Schematic of signals at opposite ends of a straw tube. $t_s$ and $t_i$ are the time shift due to different signal propagation times and the signal integration time, respectively.}
      \label{fig:time_shift}
   \end{center}
\end{figure}

Technically, this is done by integrating both signals in a narrow gate starting after the arrival of the first signal, as shown
in~\autoref{fig:time_shift}.
The signal from the far end of the tube is shifted by $t_s$ due to the longer
propagation time. Both signals are integrated during the time $t_i$.
As can be clearly seen from~\autoref{fig:time_shift}, a larger fraction is integrated for the signal arriving first.
It should be noted that $t_i$ must be larger than the maximum $t_s$, otherwise
the correlation between the coordinate and asymmetry becomes ambiguous.

\begin{figure}[htb]
   \begin{center}
      \includegraphics[width=\swidth]{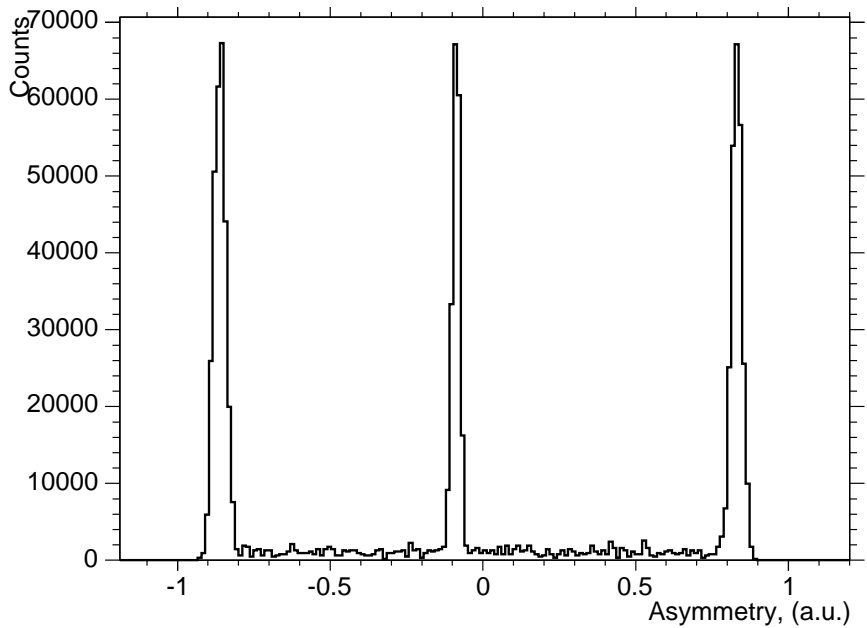}
      \caption[time-dependent signal asymmetry for the $1.5\,m$ straw.]{Time-dependent charge asymmetry distribution for an $^{55}$Fe 
                  source located at $z_{true}=\,\pm76.0$\,cm and -8.0\,cm along the $1.52$-m straw tube with the W/Re (82/18) wire of 20\,$\mu$m diameter. The 
                  integration time $t_i$ is $10\,ns$.}
      \label{fig:real_asymm}
   \end{center}
\end{figure}

A plot of the time-dependent asymmetry for the $1.52$-m straw tube
with the W/Re (82/18) wire is shown in~\autoref{fig:real_asymm} for an $^{55}$Fe source located in three places ($z_{true}=\,\pm76.0\,$cm, -8.0\,cm).
From this figure it can be seen that the amplitude of the
time-dependent asymmetry is much higher than for the linear
asymmetry, where the maximum amplitude only reaches values of about
$\pm 0.3$ for this straw tube.

\begin{figure}[htb]
   \begin{center}
      \includegraphics[width=\swidth]{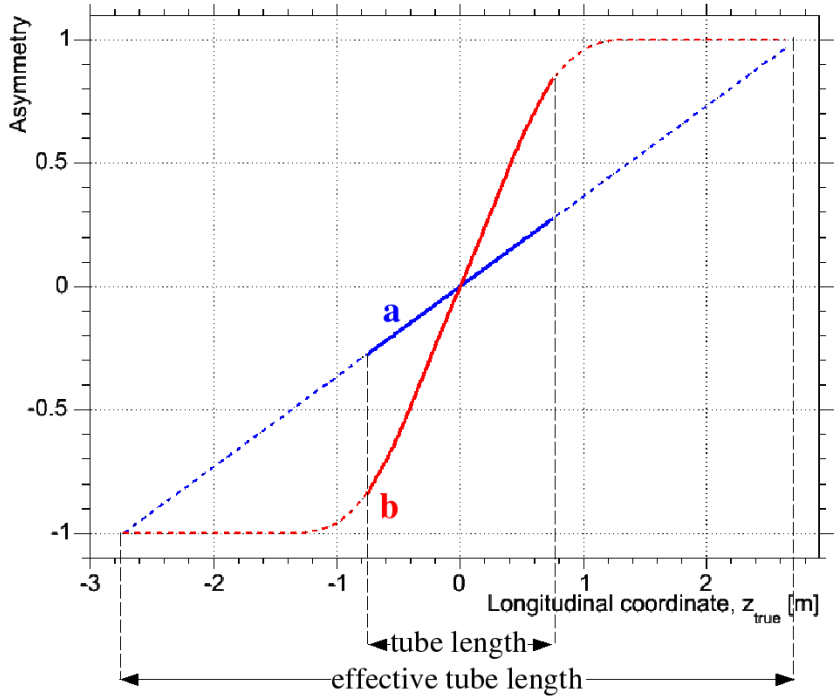}
      \caption[Comparison between the linear and time-dependent asymmetries.]{Comparison between the linear and time-dependent 
         asymmetries for the $1.52$-m straw tube with the
         high-resistance wire. The curves (a) and (b) represent the dependence of the linear
         and time-dependent signal asymmetry on the source position along the
         straw tube, respectively.
      }
      \label{fig:time_asymm}
   \end{center}
\end{figure}

Conversion of the time-dependent asymmetry into the longitudinal
position, however, requires attention because the
dependence between the asymmetry and the position is no longer linear,
as illustrated in~\autoref{fig:time_asymm}.

The time-dependent asymmetry shown in this figure (curve (b)) was
computed numerically by the following method. \autoref{equ:pss:res:sig} was used
to represent the shape of the signals from the straw tube. The
amplitudes of the signals at each end of the tube were determined for an arbitrary position of the source using the linear charge asymmetry (curve (a)
in~\autoref{fig:time_asymm}) for an effective tube length of 
546\,cm. The
arrival time difference $t_s$
was calculated for each point using the signal propagation speed
along the straw tube, which was measured to be $22.1\,$cm/ns.  Charges
obtained by integration of these signals during the integration time
$t_i$ were inserted into~\autoref{equ:assym}. The best resolution was obtained for $t_i$ equals 10\,ns. 

\begin{figure}[htb]
   \begin{center}
     \includegraphics[width=\swidth]{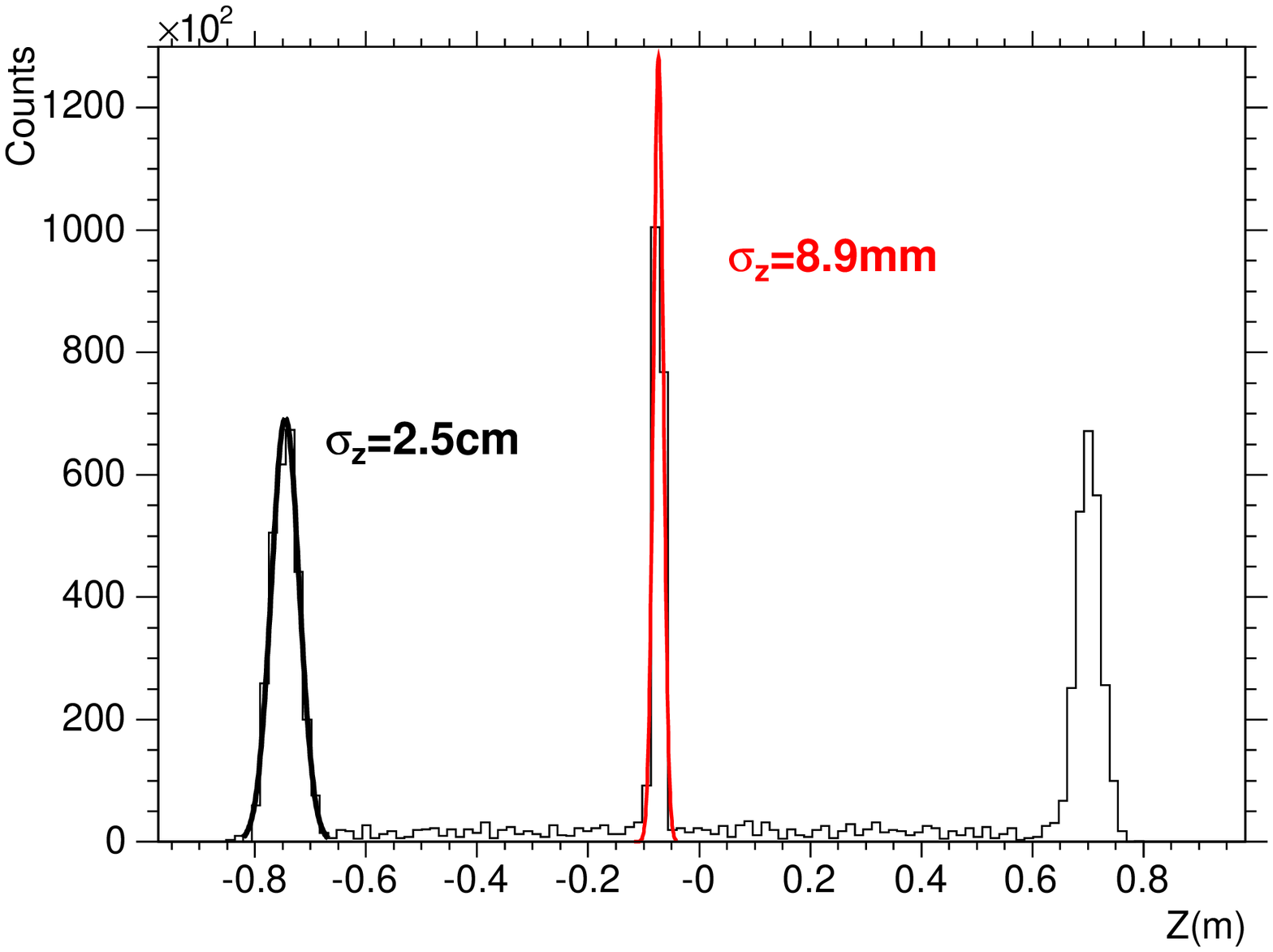}
      \caption[$z$-resolution for the $1.52\,m$ straw.]{Reconstructed longitudinal position distribution for an $^{55}$Fe source located at $z_{true}=\,\pm76.0$\,cm and -8.0\,cm along the $1.52$-m straw tube with the W/Re (82/18) wire of 20\,$\mu$m diameter, based on the time-dependent signal asymmetry. The position resolutions near the center and at the ends were found to be $\sigma=0.89\pm0.2_{sys}\,$cm and $\sigma=2.5\pm1.0_{sys}$\,cm, respectively.}
      \label{fig:real_resol}
   \end{center}
\end{figure}

The curve obtained for the time-dependent charge asymmetry was used to convert
the measured charge asymmetry (see~\autoref{fig:real_asymm}) into the longitudinal position shown 
in~\autoref{fig:real_resol}. 
Longitudinal position resolutions of $\sigma=0.89\pm0.2_{sys}\,$cm (0.6\%) and $\sigma=2.5\pm1.0_{sys}$\,cm (2.0\%) were
determined at the tube center and end, respectively. 
The systematic error increases towards the ends of the tube compared to the region near the tube center because of the non-linear dependence of 
$z_{rec}$ on the measured asymmetry.
Overall, the longitudinal position resolution is significantly improved compared to the linear charge asymmetry (cf.~\autoref{fig:long_straw}).

\begin{figure}[htb]
   \begin{center}
      \includegraphics[width=\swidth]{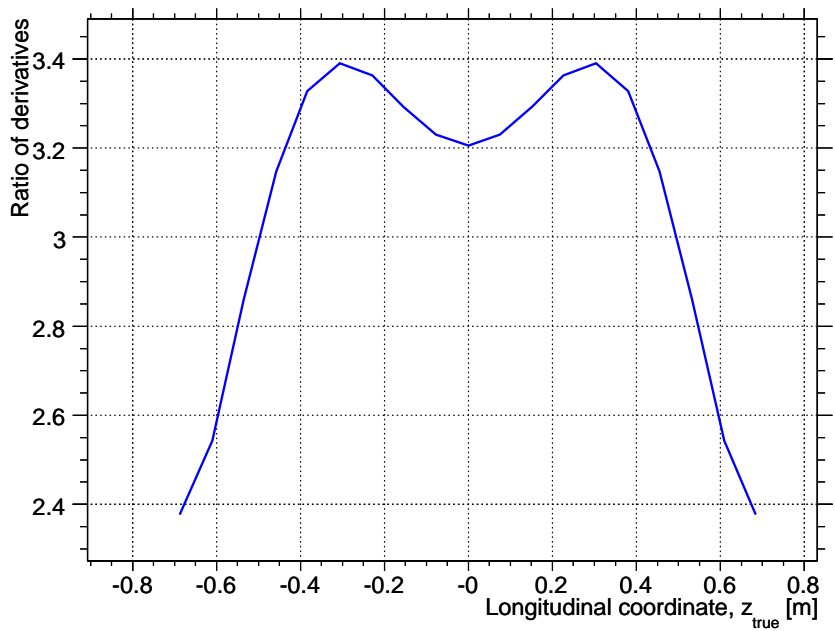}
      \caption[Ratio of derivatives of the time-dependent and linear charge asymmetries.]{Ratio of the derivatives $d\alpha/dz_{true}$ for the time-dependent asymmetry $\alpha_{td}$ and the linear charge asymmetry $\alpha_{linear}$ as a function of $z_{true}$.}
      \label{fig:ratio_asymm}
   \end{center}
\end{figure}

The longitudinal position resolution is improved in those regions where the slope of the time-dependent asymmetry as a function of $z_{true}$ is steeper than the slope of the linear method.
Since the functional dependence of the asymmetry on $z_{true}$ is non-linear (see~\autoref{fig:time_asymm}),
the improvement in the position resolution varies along the length of the straw tube.
This is illustrated graphically in~\autoref{fig:ratio_asymm}, which shows the ratio of the slopes for the time-dependent and the linear asymmetries.
The time-dependent asymmetry gives a better position resolution in the region where this ratio is greater than 1. 
Ratio of the derivatives increases for decreasing integration time, thereby improving the longitudinal position resolution.
However, at the same time both the non-linearity and the relative preamplifier noise increase, 
thereby worsening the resolution. 
Thus, the integration time is determined by optimizing the longitudinal position resolution of the straw tube.

\section{Test with $^{90}Sr$ Source}
\label{sr}
To study the applicability of this method to a realistic detector response, a collimated
$^{90}Sr/^{90}Y$ $\beta^-$ radioactive source was used to produce signals from minimum ionizing particles (MIPs). The average number of electrons 
arriving at the same moment at the anode wire is much smaller for MIP tracks compared to the clusters of ionization produced by the $^{55}Fe$ source. 
Therefore, the average signal amplitude from the straw tube is significantly smaller. In addition, the signal fluctuations are much higher due to the nature of the MIP ionization. 
This leads to a deterioration of the relative errors of the signal charge determination. To compensate these 
effects and improve the signal-to-noise ratio, the integration time $t_i$ should be increased. In this case, the position resolution was 
optimized with $t_i=$20\,ns.

\begin{figure}[htb]
   \begin{center}
      \includegraphics[width=\swidth]{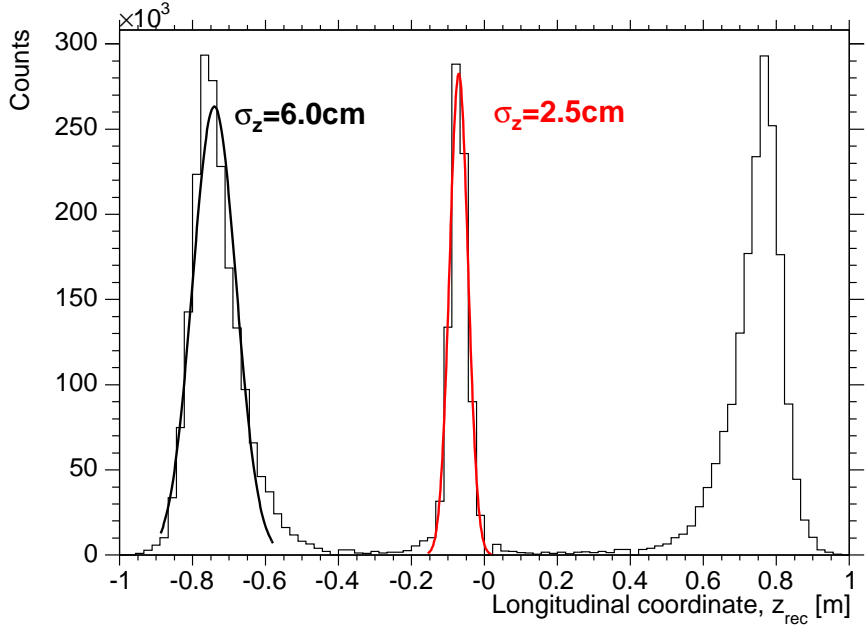}
      \caption[$z$-resolution for the $1.52\,$m straw irradiated by the $^{90}Sr/^{90}Y$ source.]{Reconstructed longitudinal position distribution for a $^{90}Sr/^{90}Y$ source located at $z_{true}=\,\pm76.0$\,cm and -8.0\,cm along the $1.52$-m straw tube with the W/Re (82/18) wire of 20\,$\mu$m diameter. The integration time $t_i$ is $20\,$ns. The position resolutions at the center and at the end were found to be $\sigma=2.5\pm1.0_{sys}$\,cm and $\sigma=6.0\pm1.2_{sys}$\,cm, respectively.}
      \label{fig:resol_Sr}
   \end{center}
\end{figure}
\autoref{fig:resol_Sr} shows the $z_{rec}$ distribution when the straw tube is irradiated at $z_{true}=\,\pm76.0\,$cm and -8.0\,cm. The observed resolutions for $z_{rec}$ near the center and at the end were found to be $\sigma=2.5\pm1.0_{sys}$\,cm (1.8\%) and $\sigma=6.0\pm1.2_{sys}$\,cm (4.0\%), respectively.

\section{Summary}
\label{sum}
Prototype straw tube detectors for the upcoming $\overline{\mbox{\sf P}}${\sf ANDA} experiment have been assembled and tested.
The longitudinal position resolution of a tube was investigated using both the linear charge division and 
the time-dependent asymmetry techniques. By adding the time information to the standard charge division method 
a significant improvement in the precision of the position reconstruction along the straw tube was demonstrated.

The longitudinal position resolution in the center and at the end of the $1.52$-m straw tube, irradiated by the $^{55}$Fe radioactive source, was determined as $1.9\,$cm (1.2\% of the tube length) and
$3.1\,$cm (2.0\%), respectively, with a maximum deviation from linearity of about 0.8\,cm (0.5\%) using the asymmetry of the fully 
integrated signal charges at both ends of the straw tube. By additionally using in the relative signal arrival time information, it was possible to improve the resolution and the linearity 
to 0.89\,cm (0.6\%) and 0.2\,mm (0.1\%) near the tube center and to 2.5\,cm (1.7\%) and 1.0\,cm (0.7\%) at the tube end, respectively.
The resolution with a $^{90}Sr/^{90}Y$ $\beta^-$ source is poorer because of the smaller signals and larger fluctuations for the minimum ionizing particles. The longitudinal position resolution in the tube center is $2.5$\,cm (1.7\%) deteriorating to $6.2$\,cm (4.2\%) at the end of the tube. The maximum deviation from linearity was about 1.2\,cm (0.8\%).

In addition to the improved resolution, the time-dependent charge asymmetry has the advantage that very short integration times can be used 
with fast shaping amplifiers, thus permitting higher rate capabilities. Also, an acceptable position resolution can be achieved even for
anode wires with relatively small resistance.

It is expected that the results obtained here can be improved since the preamplifiers used were optimized for time resolution in order 
to determine the transverse coordinate. Further studies are ongoing to optimize the signal shaping parameters of the preamplifier for both the
timing and charge integration performance.

The results obtained show good prospects for using time information to improve the precision of the charge division measurement.


\begin{thebibliography}{00}




\bibitem{Panda} M. Kotulla et al., PANDA Technical Progress Report, Darmstadt 2005,
\url{http://www.ep1.rub.de/~panda/archive/public/panda_tpr.pdf}

\bibitem{Lamina} Lamina Dielectrics Company, UK. \url{http://www.laminadielectrics.com}

\bibitem{Tof} P. Wintz et al., in AIP conf. Proc., volume 698, page 789, 2004.

\bibitem{pullia} A. Pullia et al., IEEE Trans. Nucl. Sci. 49,3269, 2002.

\end{thebibliography}
\end{document}